\documentclass[aps,prl,groupedaddress,twocolumn]{revtex4-1}
\usepackage{color}
\usepackage{xcolor}
\usepackage{epsfig}
\usepackage{graphicx}
\usepackage{amsfonts}
\usepackage[figuresright]{rotating}
\usepackage{amssymb}
\usepackage{amsmath}
\usepackage{color}
\usepackage{mathrsfs}

\usepackage{graphicx}
\DeclareGraphicsExtensions{.eps, .PNG, .jpg}

\newcommand{\ket}[1]{| #1 \rangle}
\newcommand{\bra}[1]{\langle #1 |}

\begin{document}

\title{Characterizing disordered fermion systems using the momentum-space entanglement spectrum} 
\author{Ian Mondragon-Shem, Mayukh Khan, and Taylor L. Hughes}
\affiliation{Department of Physics, University of Illinois, 1110 West Green St, Urbana IL 61801}
\date{\today}

\begin{abstract}
The use of quantum entanglement to study condensed matter systems has been flourishing in critical systems and topological phases. Additionally, using real-space entanglement entropies and entanglement spectra one can characterize localized and delocalized phases of disordered fermion systems. Here we instead propose the momentum-space entanglement spectrum as a means of characterizing disordered models. We show that localization in 1D arises from the momentum space entanglement between left and right movers and illustrate our methods using explicit models with spatially correlated disorder that exhibit phases which avoid complete Anderson localization. The momentum space entanglement spectrum clearly reveals the location of delocalized states in the energy spectrum and can be used as a signature of the phase transition between a delocalized and localized phase.
\end{abstract}

\maketitle

New light has been shed on quantum many-body ground states via their entanglement properties. Entanglement measures, such as the topological entanglement entropy\cite{aliosciatoric,wenTEE,kitaevTEE,fradkinleigh,flammia} and the entanglement spectrum\cite{lihaldane,Thomale2010A,bernevigarovasthomale,fidkowski2010,Pollmann2010,hughesinversion,turner2010}, with their origin in quantum information theory have begun to uncover subtle features of topological insulators and systems with topological order. One field where entanglement characteristics have begun to make progress is in disordered materials. For example, Ref. \onlinecite{chakravarty1,2} showed a connection between the position-space entanglement entropy and a multi-fractal scaling exponent which was extended by Ref. \onlinecite{xiaochen} to establish a connection between R\'enyi entropies and the universal part of the multi-fractal spectrum. Additionally Ref. \onlinecite{prodanhughesbernevig,gilbertbernevighughes} showed that the level-spacing statistics of the entanglement spectrum match the same statistical ensemble as the energy spectrum. Since entanglement  in position-space captures information about long-range correlations it is natural that it should reveal information about localized and de-localized states.

In this article we also consider the entanglement characterization of disordered fermion models, but from the perspective of momentum space instead of position space. The advantage (and from another perspective the disadvantage) of entanglement is that it crucially depends on the way Hilbert space is partitioned into two sectors. The partitioning can be done in \emph{any} way, though typically the partition (or cut for short) is made in position-space. While the dependence on a cut makes entanglement non-unique, the benefit is that there is a possibility to learn more information about a system by taking different cuts. The notion of cutting the system in momentum space was studied for translationally invariant 1D spin-chains in Ref. \onlinecite{bernevigarovasthomale}, but, as we will mention again below, this type of cut yields no information at all for translationally-invariant, non-interacting fermion systems. While a momentum-space cut is useless when translation symmetry is preserved, we show, however,  that it does lead to valuable information if the system is disordered. The disorder couples/entangles different momentum states through scattering and the momentum space entanglement spectrum shows clean signatures to characterize regions of localized and de-localized states. In this work we focus solely on 1D models where the essential idea is that right-moving states, which were delocalized in the translationally invariant limit, remain delocalized if the disorder does not entangle them with left-movers.

\begin{figure}
\begin{center}
\includegraphics[scale=0.063]{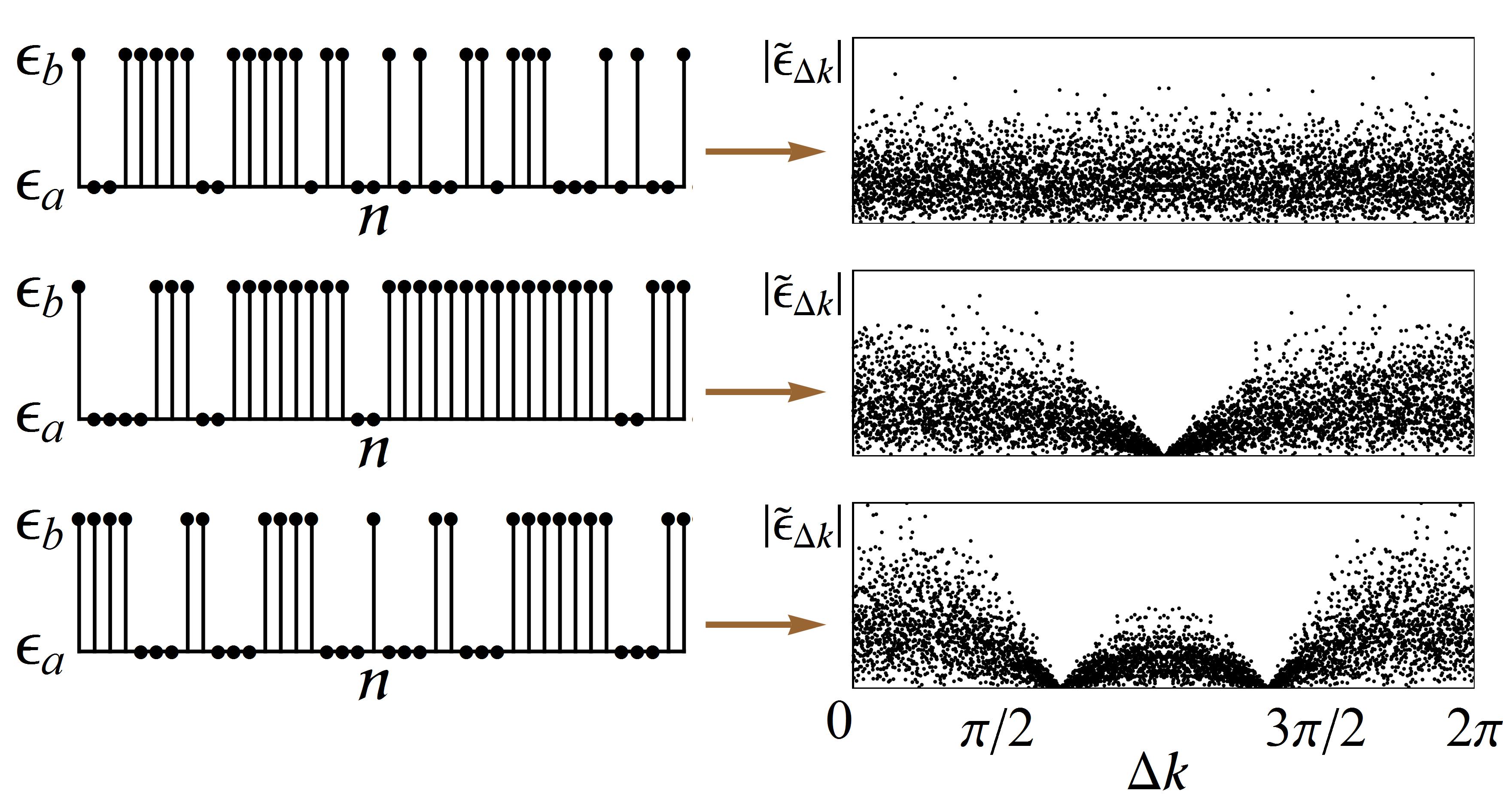}
\caption{Profile of position (left) and Fourier (right) distributions of the $n$-mer disorder potential, with $n=1,2,3$ (top to bottom).}\label{MomDist}
\end{center}
\end{figure}

To illustrate localization transitions in 1D we must use special classes of models as it is well known that disordered, 1D electronic systems become insulating due to the phenomenon of Anderson localization. An exception to Anderson localization in 1D was discovered by Dunlap \textit{et al.} who predicted that a model with \emph{spatially correlated} disorder, the so-called random dimer model (RDM), could avoid complete localization (at least in finite-size systems)\cite{Dunlap1990}. 
The RDM model is described by a tight-binding lattice Hamiltonian
\begin{equation}
H= \sum_{n=1}^{N} t \left( c^{\dagger}_{n+1}c_{n}+ c^{\dagger}_{n}c_{n+1}\right)+\sum_{n=1}^{N} \epsilon_n c_n^{\dagger} c_n, \label{RD}
\end{equation}
where $n$ labels the $N$ sites of the lattice, $t$ is the nearest neighbor hopping , and $\epsilon_n$ represents a site-dependent energy. In the RDM the site energies, which represent the random disorder have a specific structure: (i) $\epsilon_n$ can take two possible values $\epsilon_a$ or $\epsilon_b$ (ii) the two energies are randomly placed on each site with the constraint that when $\epsilon_a$ is chosen it always placed on two consecutive sites (thus a dimer, see Fig. \ref{MomDist}). This constraint is what renders the disorder spatially correlated.  Dunlap \textit{et al.}  showed that the RDM model possesses a pair of delocalized eigenstates that can propagate throughout the lattice without being localized by the disorder. These states are degenerate and occur at an energy $E=\epsilon_a,$ which is termed the resonant energy.  An additional $\mathcal{O}(\sqrt{N})$ states in the neighborhood of the resonant energy remain extended as well, and as a result, a finite fraction of the eigenstates are able to resist Anderson localization. Thus, a measurable conductance can in principle be observed in finite-length wires or polymers\cite{philipsScience}.

The RDM can be further generalized to the ``$n$-mer" case in which chains of length $n$ and energy $\epsilon_a$ are randomly placed throughout the lattice; the $n=1$ case corresponds to uncorrelated disorder, and $n=2$ to the RDM. The general case has $n-1$ resonant energies at $E_n(m)=\epsilon_a-2t\cos\left(\pi m/n\right),\, m=1,\cdots, n-1$.
In what follows, we will mainly focus on the $n=2$ dimer and $n=3$ trimer (RTM) cases. Without loss of generality we set $\epsilon_a=0$ so that the resonance  energy $E_2=0$ ($E_3=\pm t$) for the dimer (trimer) model and the disorder strength is quantified by $\epsilon_b-\epsilon_a=\epsilon_b.$ For the RDM the delocalized states exist for $\vert\epsilon_b\vert<2t$ and for the RTM delocalized states at $E_3=-t (+t)$ for $-3t<\epsilon_b<t (-t<\epsilon_b<3t).$

\begin{figure}
\begin{center}

\includegraphics[scale=0.0415]{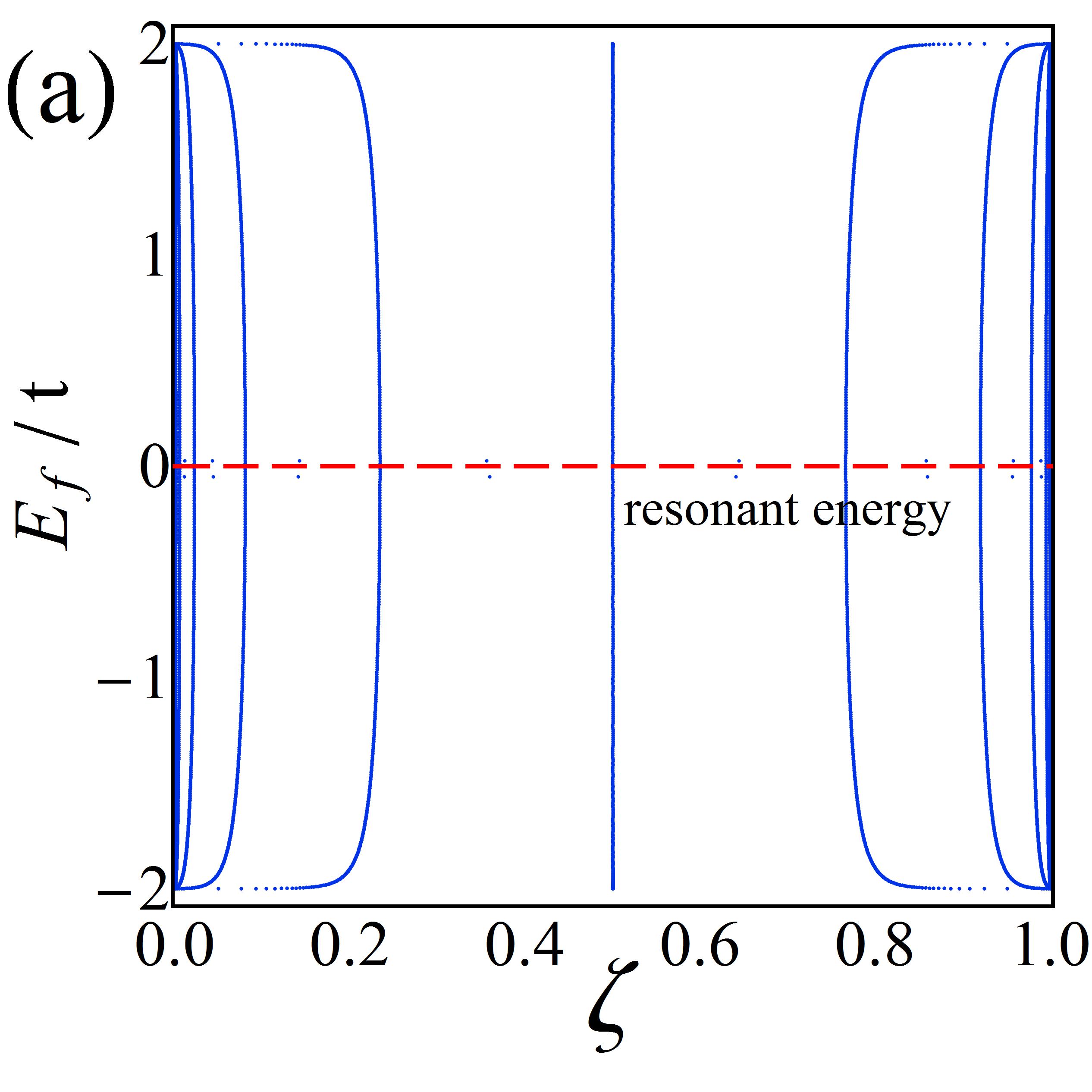}
\includegraphics[scale=0.0415]{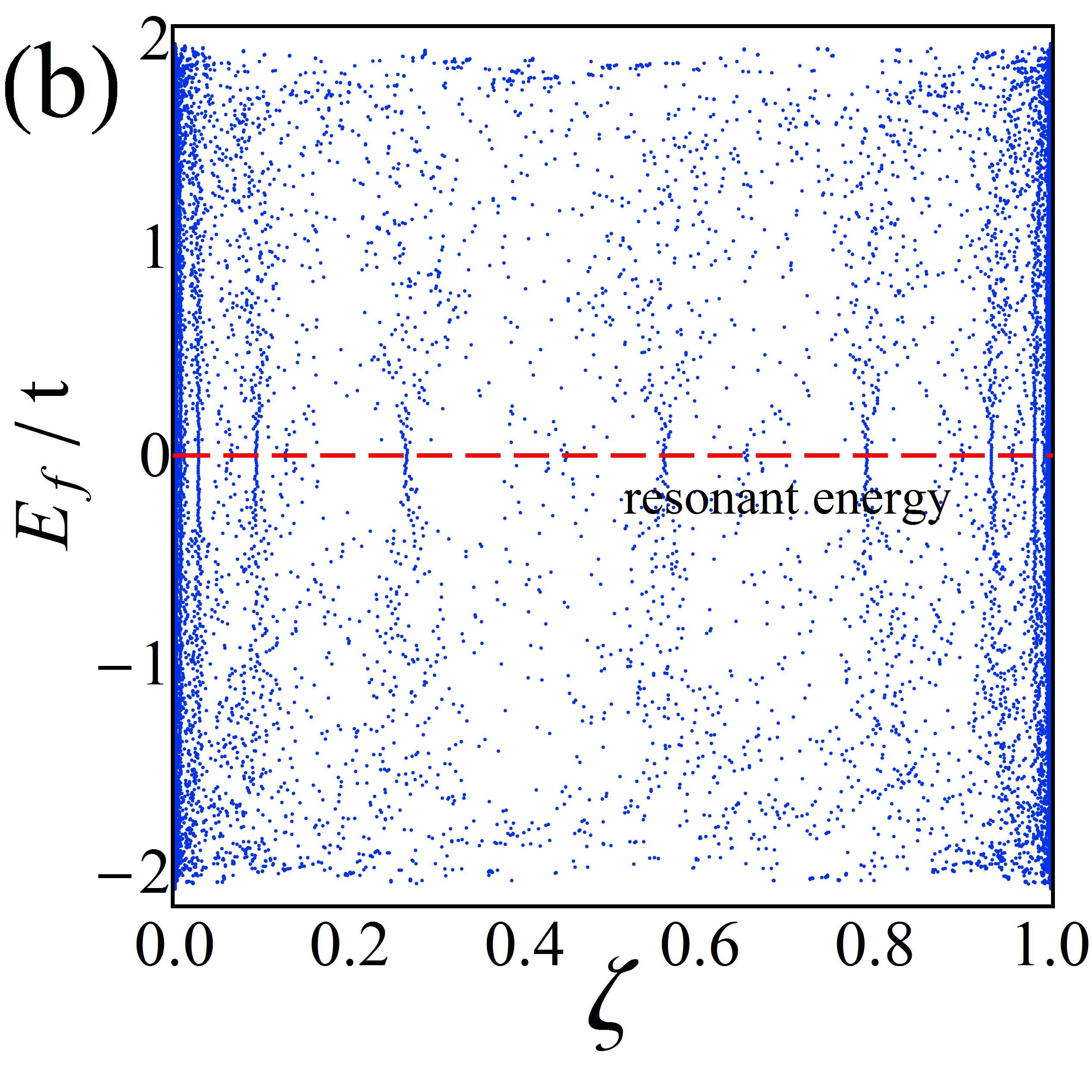}

\includegraphics[scale=0.040]{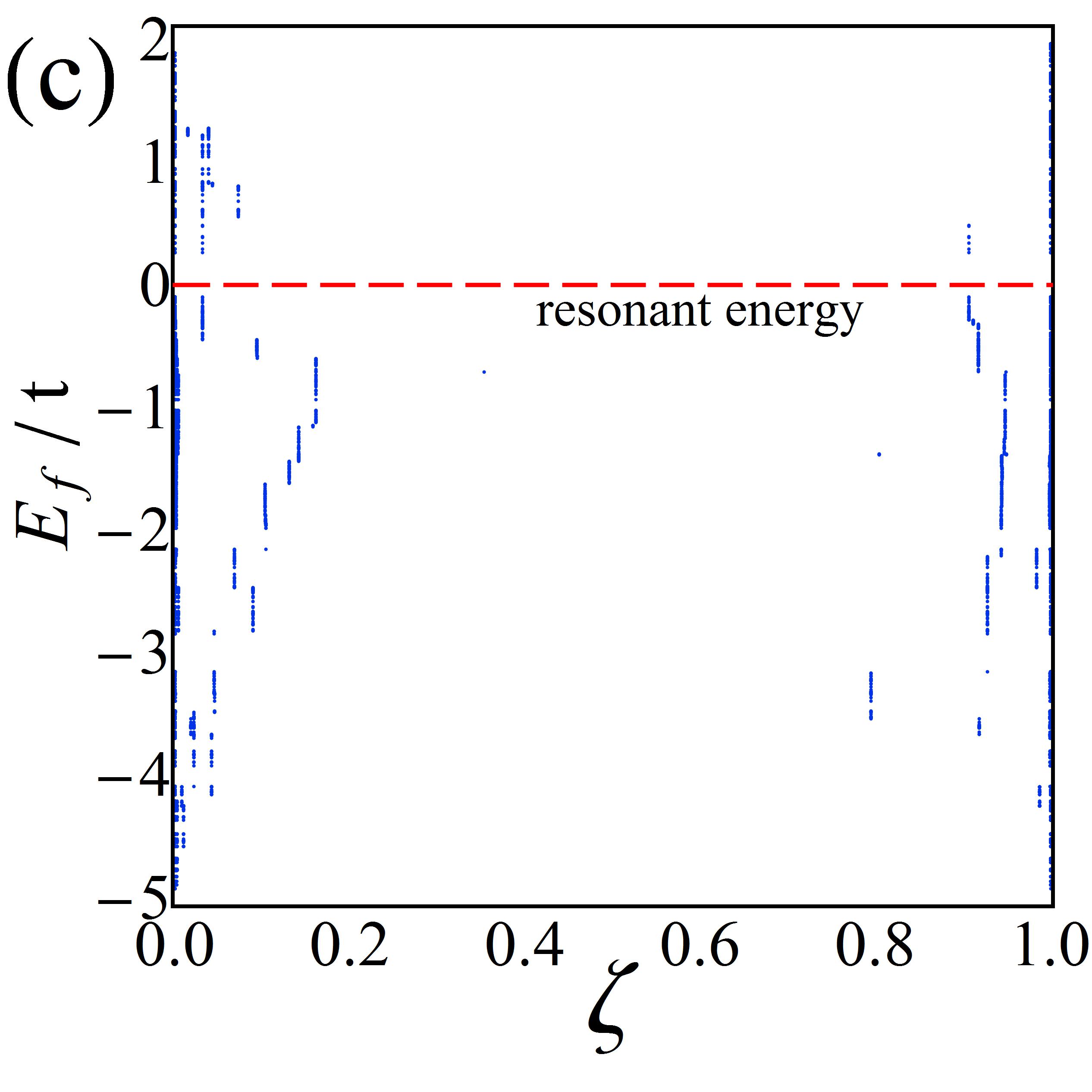}
\includegraphics[scale=0.054]{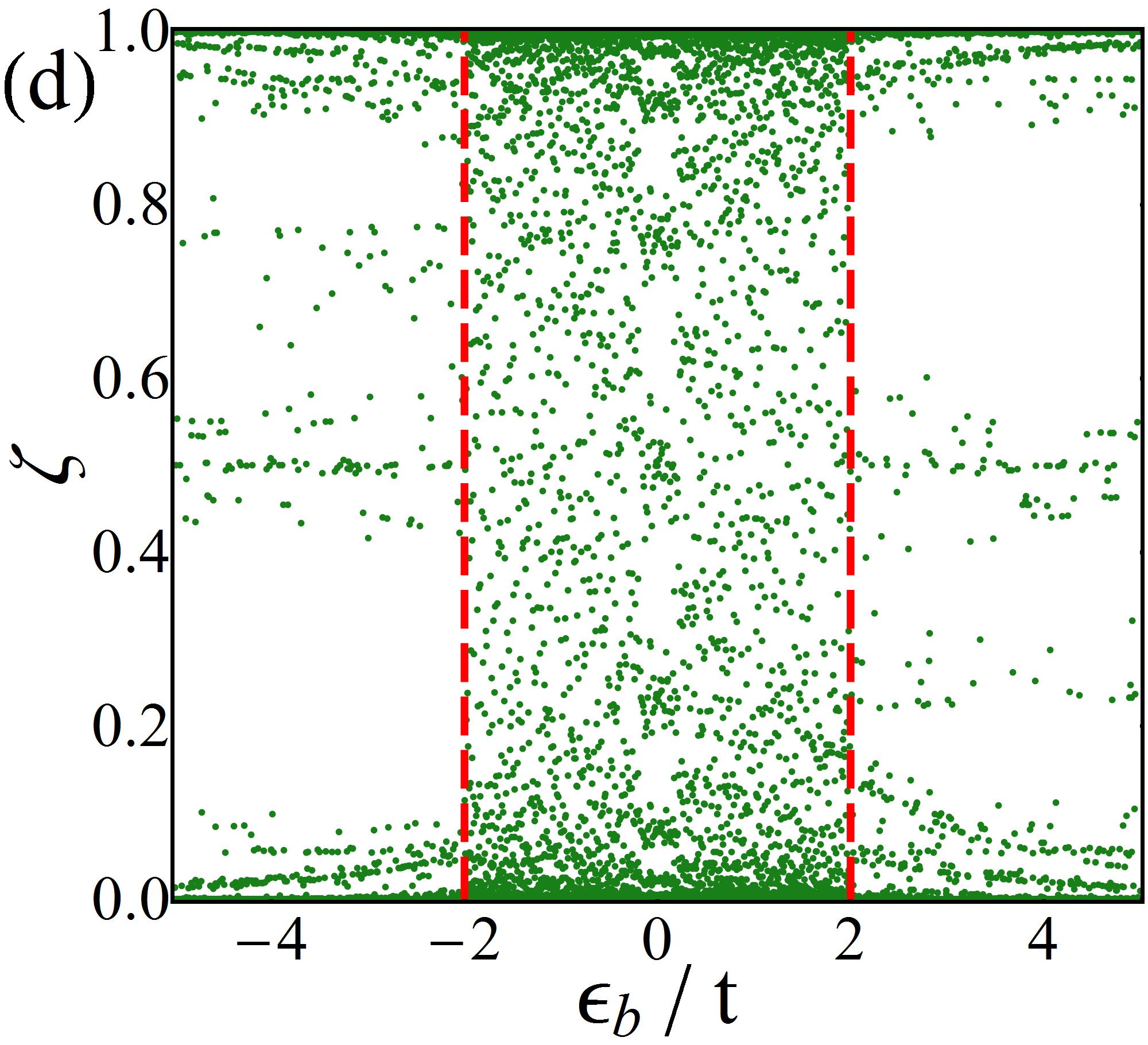}

\caption{ Entanglement spectrum as the Fermi energy is varied, with (a)$\epsilon_b=0.0 t,$ (b) $-0.1t,$ and (c)$-3.0 t.$ The red dashed lines denote the resonant energy. (d) Entanglement spectrum as the disorder strength is varied, with $E_f=0$ fixed at the resonant energy. The red dashed lines denote the critical point beyond which the extended states of the random dimer model are lost.}\label{SpatialEnt}
\end{center}
\end{figure}

In order to make contact with previous work, and for comparison to our main results, we will first consider the spatial entanglement of the RDM. We divide the Hilbert space into two subspaces $\mathcal{A}$ and $\mathcal{B}$, which contain sites confined to the region $n \in [1,N/2]$ and $n \in [N/2+1,N]$, respectively. We will focus on the single-particle entanglement spectrum calculated via Peschel's method which considers the eigenvalues  $\zeta_i$ of the correlation matrix  $C_{ij}=\bra{\Omega}c^{\dagger}_i c^{\phantom{\dagger}}_{j}\ket{\Omega}$ where $\ket{\Omega}$ is a free-fermion ground state and  $i,j \in [1,N/2]$ \cite{peschelreduced}. The $\{\zeta_i\}$  form the single-particle entanglement spectrum and always lie between $0$ and $1$ with eigenvalues near $1/2$ yielding the most entanglement.
When we calculate momentum-space entanglement later the only difference is that $i,j$ in $C_{ij}$ will refer to different single-particle momenta, which are discrete indices because $N$ is finite, and they will be restricted to lie in certain regions of momentum space instead of position space.

In Figs. \ref{SpatialEnt}a,b,c we show the entanglement spectrum as a function of the Fermi energy for disorder strengths fixed at the values $\epsilon_b=0.0 t,$ $-0.1t,$ and $-3.0 t$ respectively (note that only one disorder realization is shown and no averaging has been performed). 
For the clean system we see large entanglement for all values of the Fermi-level within the band. For $\epsilon_b=-0.1t$ we still see a lot of entanglement and one can see by eye that there is level-repulsion in the entanglement spectrum for Fermi-energies in a window around the resonant energy $E=0.$ This is reminiscent of the level repulsion commonly seen in the \emph{energy} spectrum for energy ranges containing delocalized states, and which was confirmed to exist in the entanglement spectrum in Ref. \cite{prodanhughesbernevig}. Finally, for $\epsilon_b=-3.0t$, which is tuned to the localized insulating phase, we see that the entanglement is suppressed compared to the other figures with  eigenvalues clustered near $\zeta=0$ and $\zeta=1$ which heuristically signals the absence of delocalized states. In Fig. \ref{SpatialEnt}d, we show the entanglement spectrum as a function of disorder, with the Fermi level tuned exactly to the resonance energy. Spatial entanglement appears to be more pronounced  when $\vert \epsilon_b \vert \le 2t$, as expected in the delocalized phase. There is a somewhat clear distinction between the localized and delocalized phases based on the density of entanglement eigenvalues away from $\zeta=0,1.$ In fact, even the highly entangled modes in the localized phase are most likely due to local entanglement \emph{i.e.} \emph{localized} states near the cut which happen to have weight on both sides of the partition.

While some features of the phases of the RDM are apparent from the spatial entanglement, we will show that performing a momentum-space cut leads to more explicit results. A trivial, but ultimately illustrative example is to consider a momentum space cut in the translationally invariant limit. In this limit, since momentum is conserved, entanglement is completely absent and the set of $\{\zeta_i\}$ will have a number of 1 (0) eigenvalues equal to the number of occupied (unoccupied) single-particle momentum states which depends on the Femi level. Thus, for a clean system, where all states can be considered as delocalized plane-waves, there is no entanglement. Once disorder is turned on, momentum states are mixed and the $\{\zeta_i\}$ will have more generic eigenvalues, but the key to locating regions of extended states is to search for Fermi-levels where the entanglement is \emph{suppressed}. This is the opposite of the spatial entanglement paradigm. The other important concept is that the clearest signatures are seen if the momentum cut is chosen so that region $\mathcal{A}$ consists of all the right-movers and $\mathcal{B}$ contains all the left-movers. An essential ingredient for determining the cut is thus the translationally invariant bandstructure which, as we will see in detail later, is crucial for determining the the single-particle group velocities and thus the best momentum cut.

\begin{figure}
\begin{center}
\includegraphics[scale=0.0415]{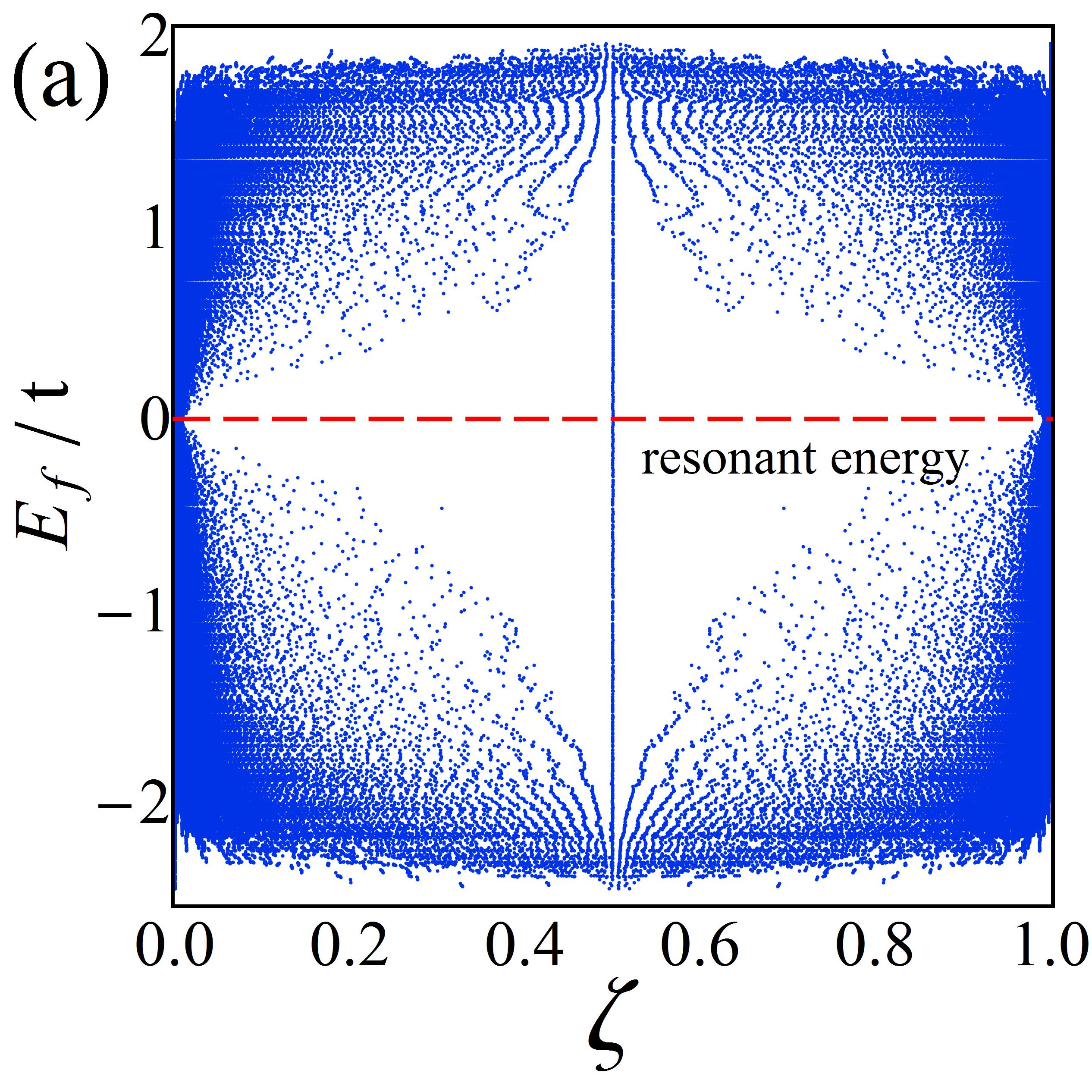}
\includegraphics[scale=0.0415]{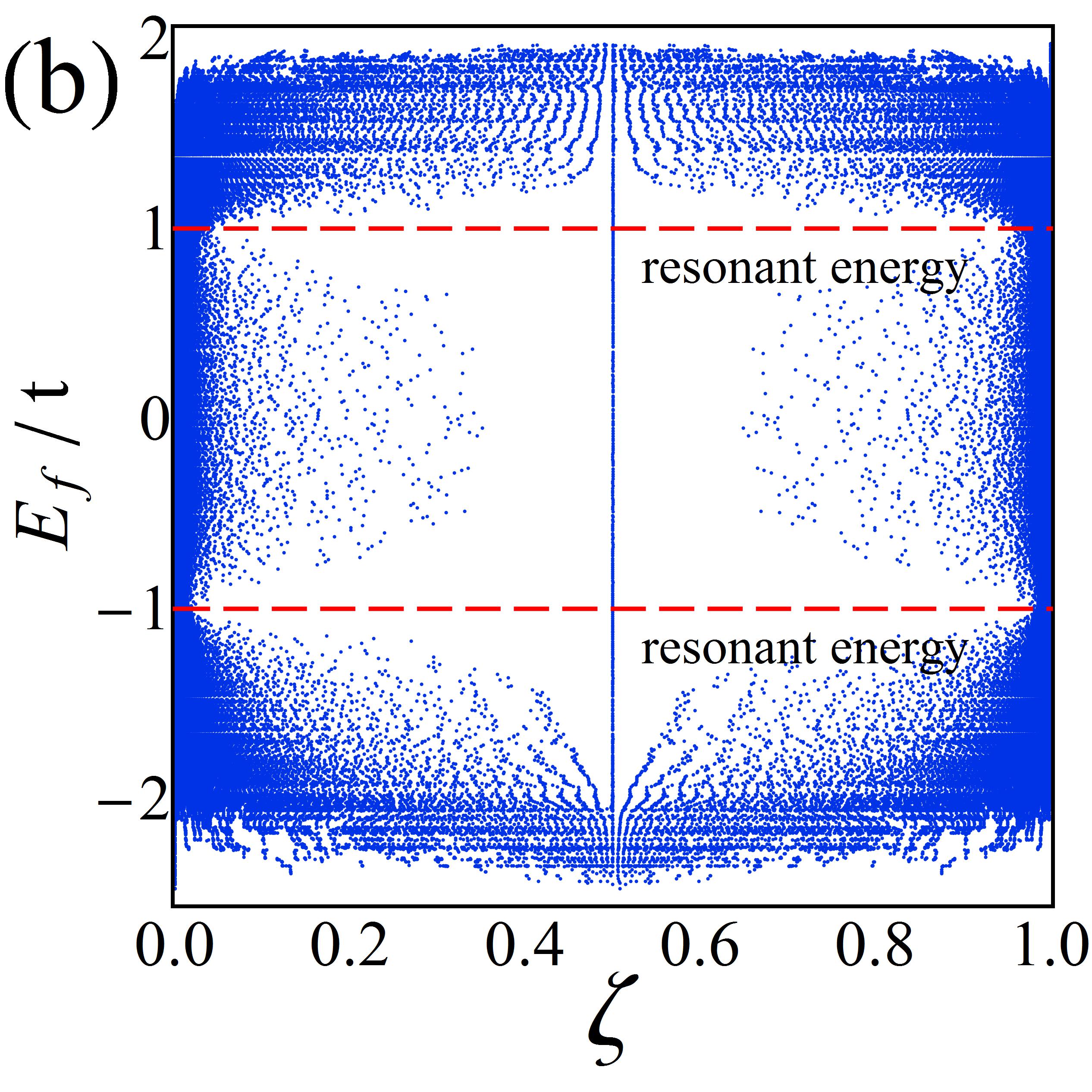}

\includegraphics[scale=0.053]{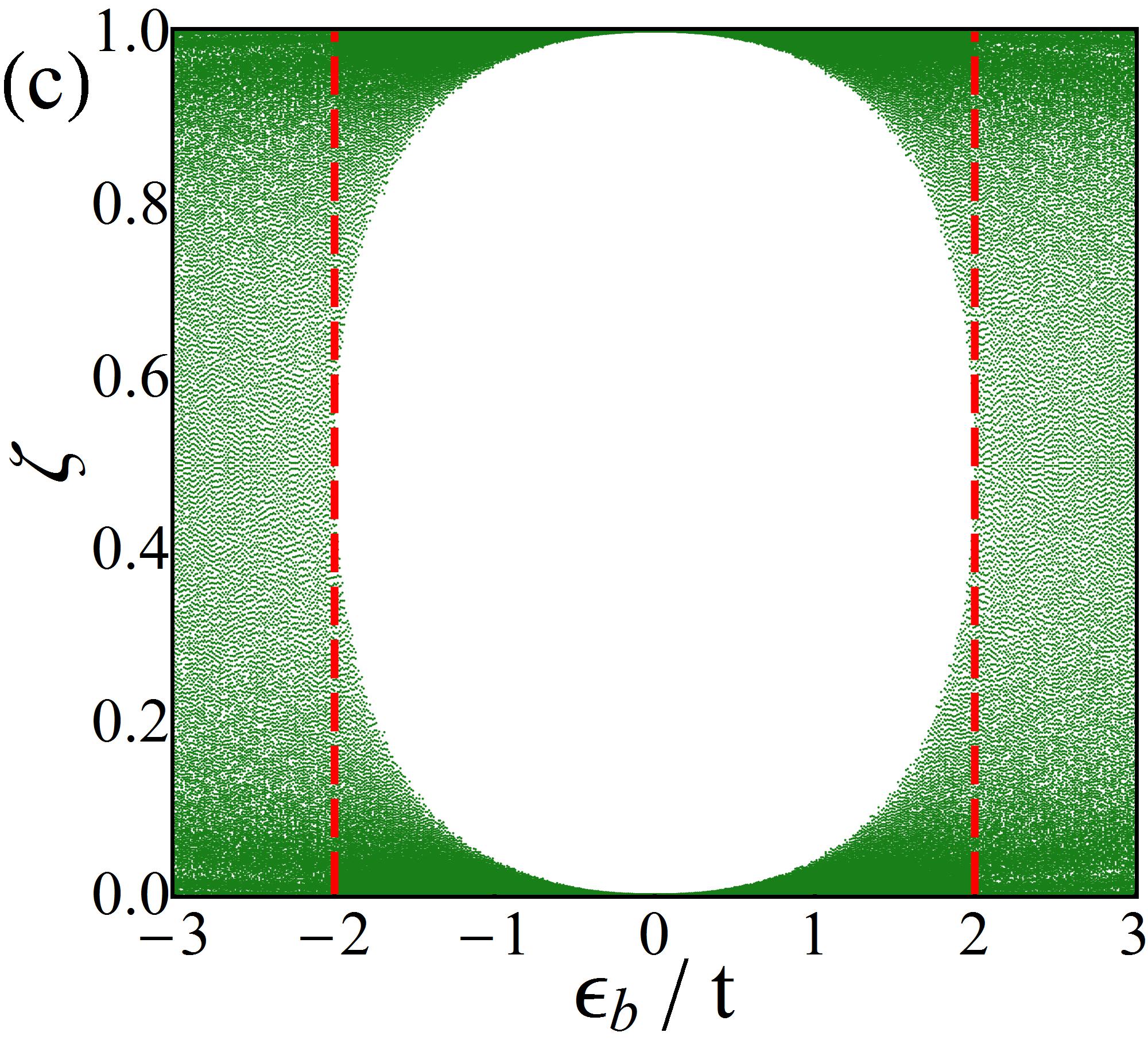}
\includegraphics[scale=0.053]{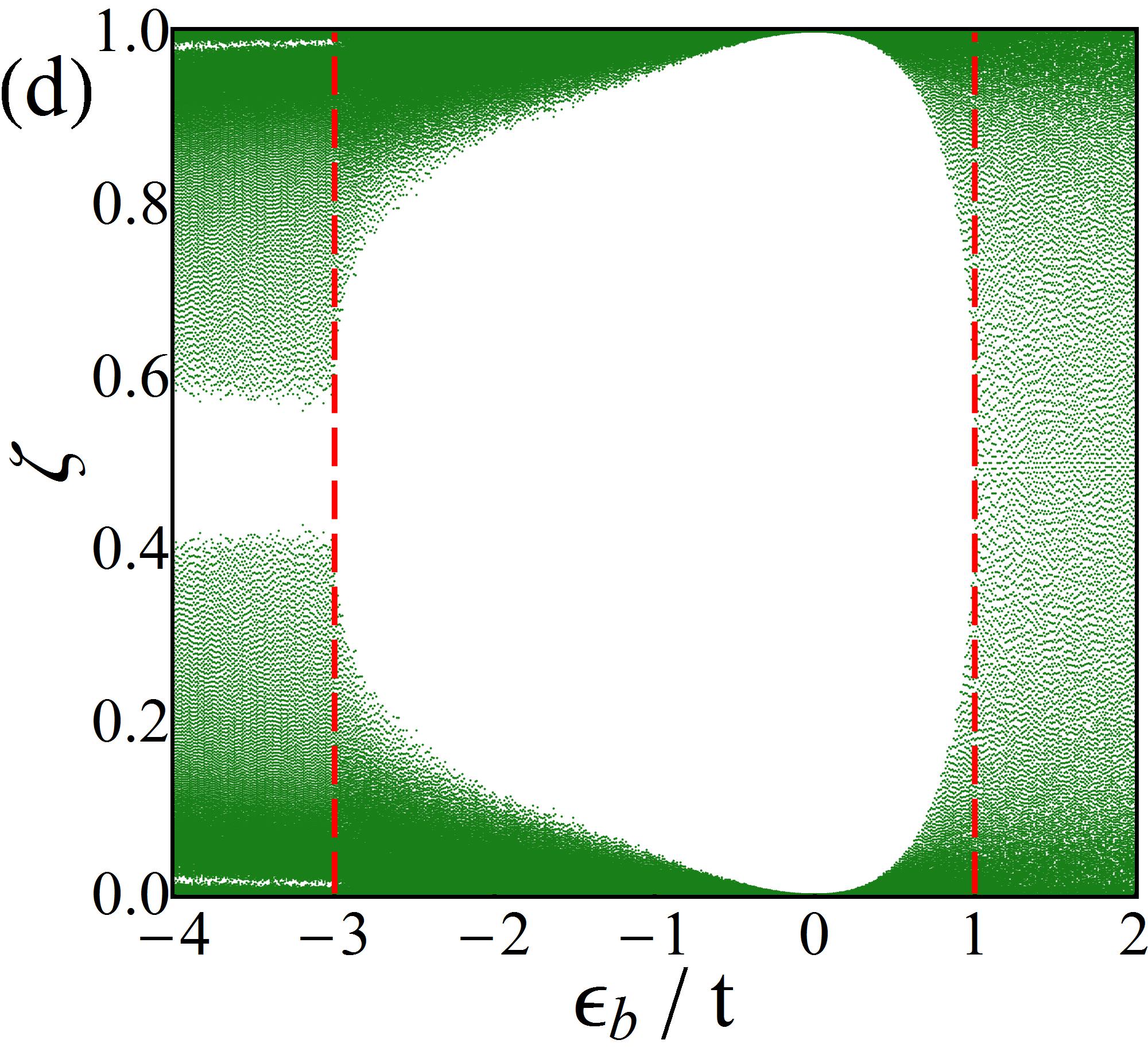}
\caption{ Entanglement spectrum as the Fermi energy is varied for (a) disordered dimers  (b) disordered trimers  for $\epsilon_b  = -0.5  t$.  Entanglement spectrum as the disorder strength is varied, for (c) disordered dimers (d) and disordered trimers  for $E_f$ at resonance.}\label{EntMom}
\end{center}
\end{figure}

For the RDM and RTM the relevant partitioned subspaces have momenta restricted so that $k_{\mathcal{A}}\in [0, \pi]$ and $k_{\mathcal{B}}\in [\pi, 2\pi]$ since these regions make up the left and right movers respectively.   In Figs. \ref{EntMom}a,b  we show the resulting entanglement spectrum for the RDM and  RTM respectively  as a function of the Fermi level, for this choice of partition. The disorder strength was set at $\epsilon_b=-0.5t$ for both the RDM and RTM. 
Remarkably, the resonant energies are revealed with great clarity where a marked suppression of entanglement occurs. In the vicinity of the resonant energies $\mathcal{O}(\sqrt{N})$ of the $\zeta_i$ approach either $0$ or $1$. On the other hand, for Fermi energies away from the resonances, the spectrum fills up with entangled eigenvalues due to scattering between left-movers and right-movers. In Figs. \ref{EntMom}c,d we show the entanglement  spectrum as a function of disorder strength for the RDM at Fermi-level $E_F=E_2(1)=0$ and the RTM at Fermi-level $E_F=E_3(1)=-t$ respectively.
It is clear from Fig. \ref{EntMom}c that the entanglement is strongly suppressed in the RDM up to the phase transition at $\vert \epsilon_b \vert=2t$, where the extended states are lost. Similarly, for the RTM we see in Fig. \ref{EntMom}d that the state at $E_3(1)=-t$ only remains delocalized for $-3t<\epsilon_b<t.$ One other feature of note is the suppressed entanglement in the RTM for $\epsilon_b<-3t.$ This effect is not generic and is due to a pathology of $n$-mer models (for $n>2$) in the strong disorder regime, which is dominated by bound states localized on the $n$-mers themselves near the resonant energies. The position of the resonant Fermi-level with respect to the energy of the $n$ discrete modes on each $n$-mer leads to this spurious suppression which we will not discuss any further. 

To understand this behavior of the RDM we study the Hamiltonian in the momentum-space basis replacing $c_n=\frac{1}{\sqrt{N}}\sum_{k}e^{ik n} c_k$ in Eq.(\ref{RD}). This leads to the expression 
\begin{equation}
H=\sum_{k}\mathcal{E}(k) c^{\dagger}_k c_k+\sum_{k,\,k'} \tilde{\epsilon}_{\Delta k} c_{k}^{\dagger} c_{k'},
\end{equation}
where $\mathcal{E}(k)=\epsilon_a+2t\cos k$. The matrix elements $\tilde{\epsilon}_{\Delta k}$ of the disorder potential read
$\tilde{\epsilon}_{\Delta k} =(\epsilon_b-\epsilon_a)\left[\delta_{k,k'}- \, f\left({\Delta k},\{r_i\}\right) S(\Delta k) \right],$
where
$f(\Delta k, \{r_i\})= \frac{1}{\sqrt{N}}\sum_{i} e^{i r_i (k-k')},$ and $S(\Delta k)=\frac{1}{\sqrt{N}}\left( 1+e^{i(k-k')}\right).$
Here, $\Delta k=k'-k,$  the $r_i$ denote the random positions of the dimers, and we have restored $\epsilon_a$ to show the dependence explicitly. The function $f\left({\Delta k},\{r_i\}\right)$ captures the random part of the disorder since it depends explicitly on the positions of the dimers whereas  the structure-factor $S(\Delta k)$  is a \emph{coherent} sum of phases, independent 
of any randomness. The function $S(\Delta k)\approx \frac{i}{\sqrt{N}}\left(\Delta k-\mathcal{Q}\right)+\cdots,$ when expanded around $ \mathcal{Q}=\pi$, which, even after convolution with $f(\Delta k,\{r_i\}),$ suppressed the scattering events $k \rightarrow k' + \mathcal{Q}$ as shown in Fig. \ref{MomDist}. The generalization to the arbitrary $n$-mer case can readily be made by replacing the structure factor with $S_n(\Delta k)=\frac{1}{\sqrt{N}}\sum_{m=0}^{n-1} e^{im(k-k')}$  which has zeroes at $\mathcal{Q}_n(m)=2\pi m/n,$  $ m=1,\cdots,n-1$. In Fig. \ref{MomDist}, we show examples of the disorder distribution in both position and momentum space respectively, for the $n=1,2,3$ cases.

 For generic weak disorder, scattering predominantly mixes degenerate single-particle momentum states $\ket{q_{1,2}}.$ If the $\ket{q_{1,2}}$ have opposite group velocity then this mixing leads to backscattering, localization, and entanglement between left-movers and right-movers. 
 If we consider the RDM model, then states satisfying $q_{1}-q_2= \mathcal{Q}$ are never mixed. A resultant suppression of momentum entanglement occurs when this condition is met by degenerate states with opposite group velocity.  For the RDM the momenta at which the condition $\mathcal{E}(q_{1})=\mathcal{E}(q_2)=\mathcal{E}(q_1-\mathcal{Q})$ is satisfied are given by $q_{1,2}=\pm\pi/2.$ At these momenta the resonant energy for the RDM is $E_{2}(1)=\mathcal{E}\left(q_1\right)=\epsilon_a+2 t\cos\left(\pi/2\right)=\epsilon_a$ which is exactly what we noted earlier, though we chose $\epsilon_a=0$ for convenience.
It then follows that the suppression of left and right-mover entanglement matches the existence of extended states in the RDM model near the resonant energy. This connection constitutes the main result of this work.
By analogy with the dimer case, the resonant energies of the $n$-mer model are determined by satisfying $\mathcal{E}(q)=\mathcal{E}(q-\mathcal{Q}_{n}(m)).$ Thus, we see that, whenever $E_F$ is at one of the resonant energies, the absence of left and right-moving entanglement is correlated with the presence of extended states in the system, just like in the particular $n=2$ dimer case.

\begin{figure}
\begin{center}
\includegraphics[scale=0.045]{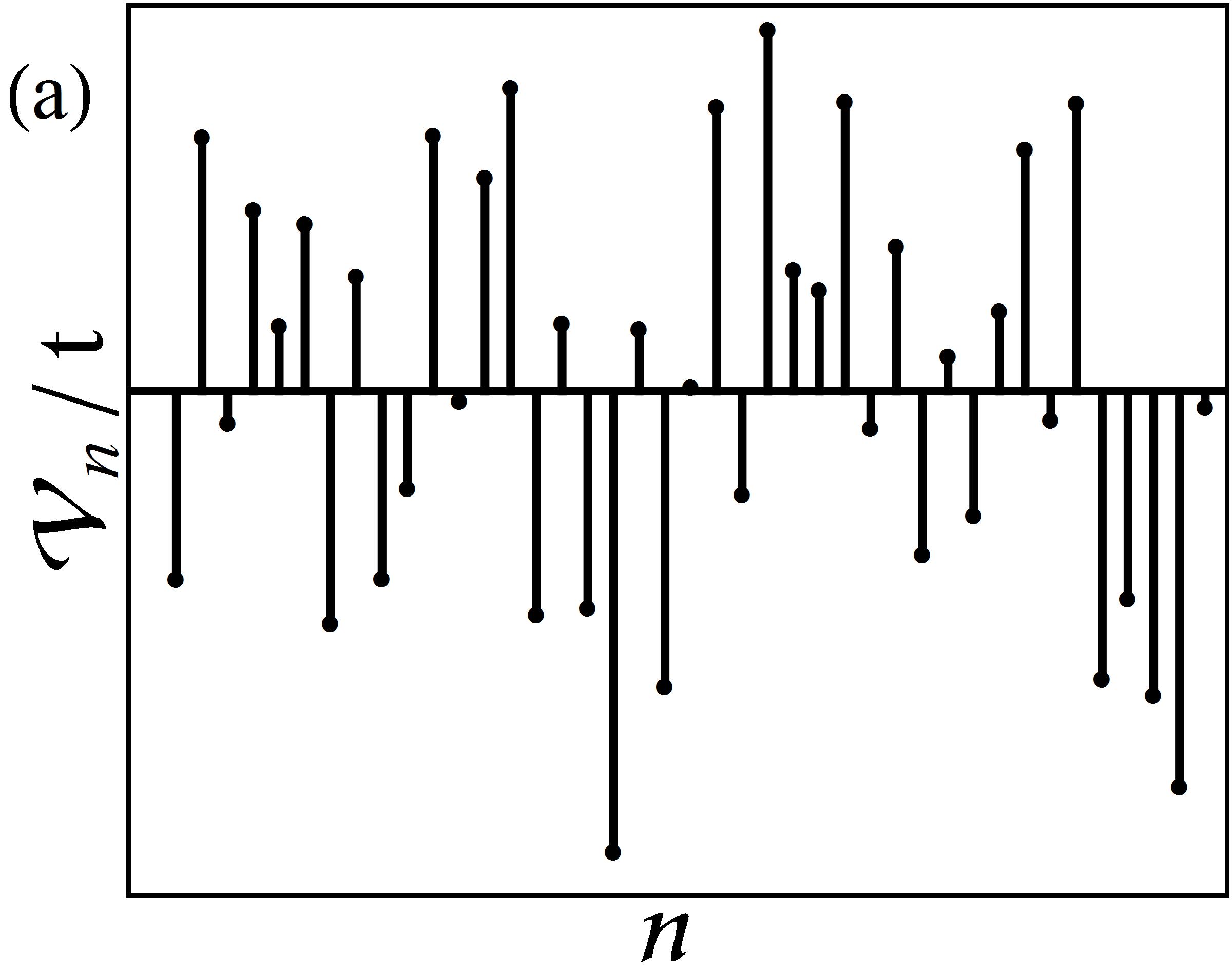}
\includegraphics[scale=0.04]{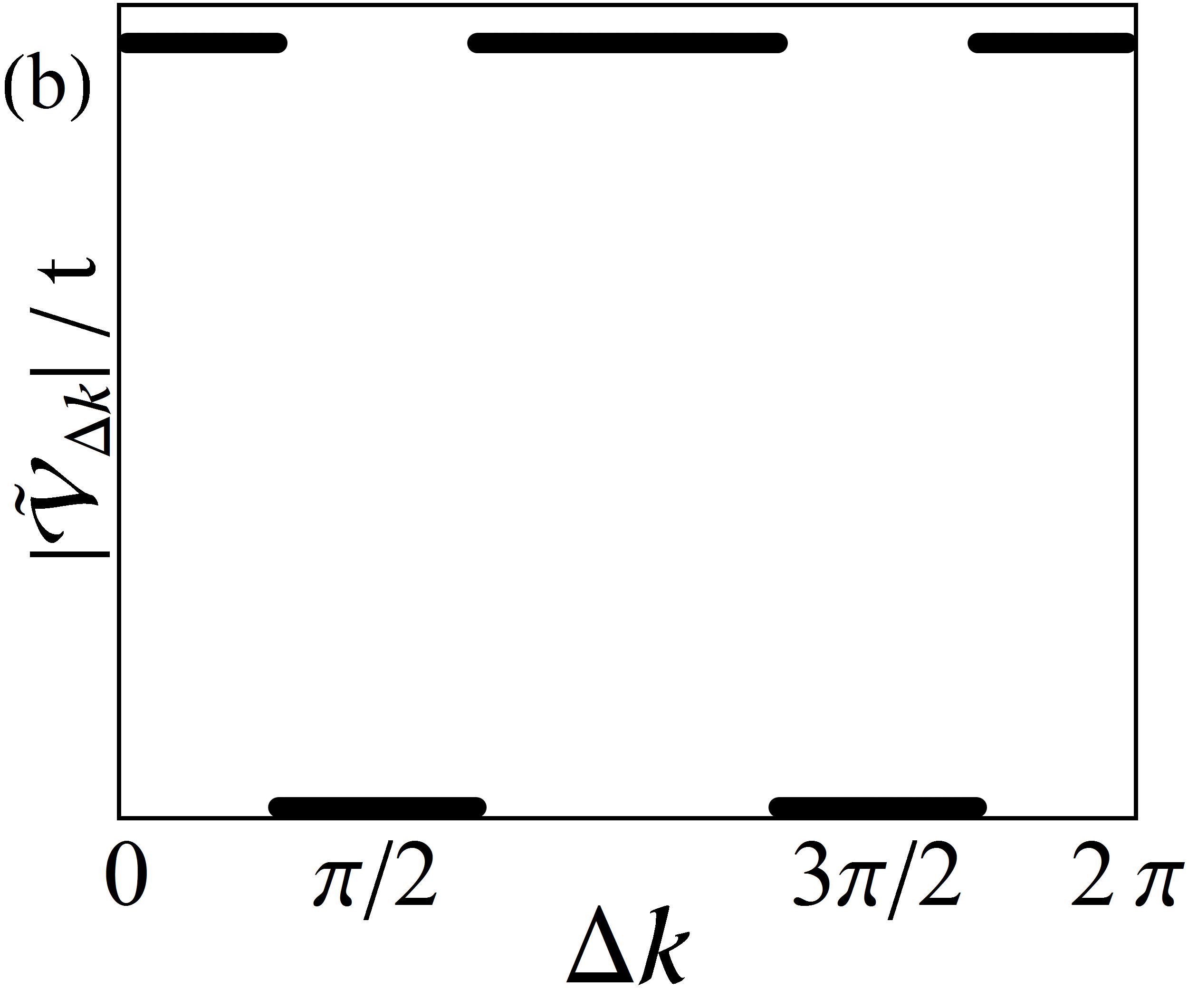}
\vspace{0.35cm}
\includegraphics[scale=0.042]{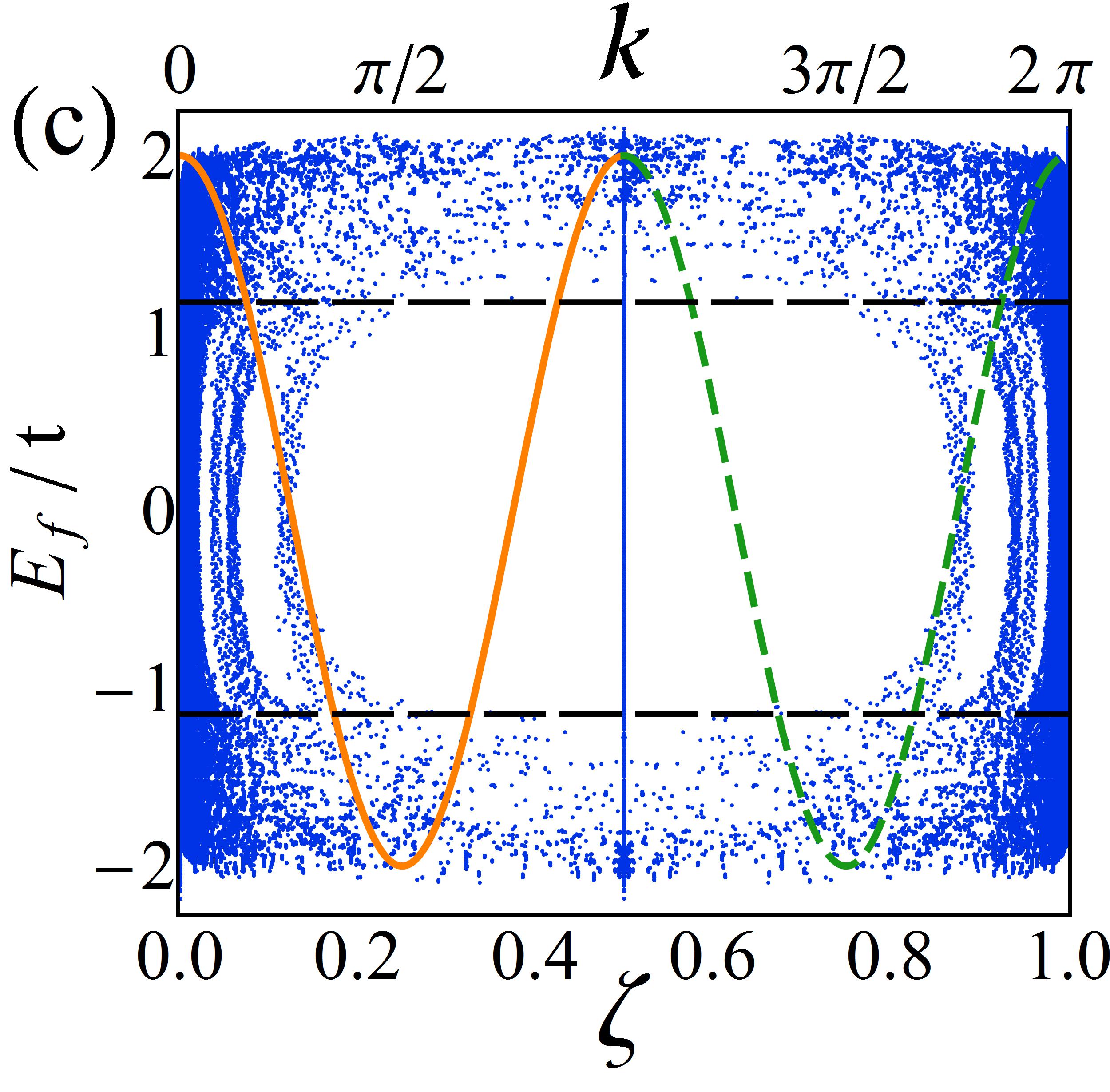}
\includegraphics[scale=0.042]{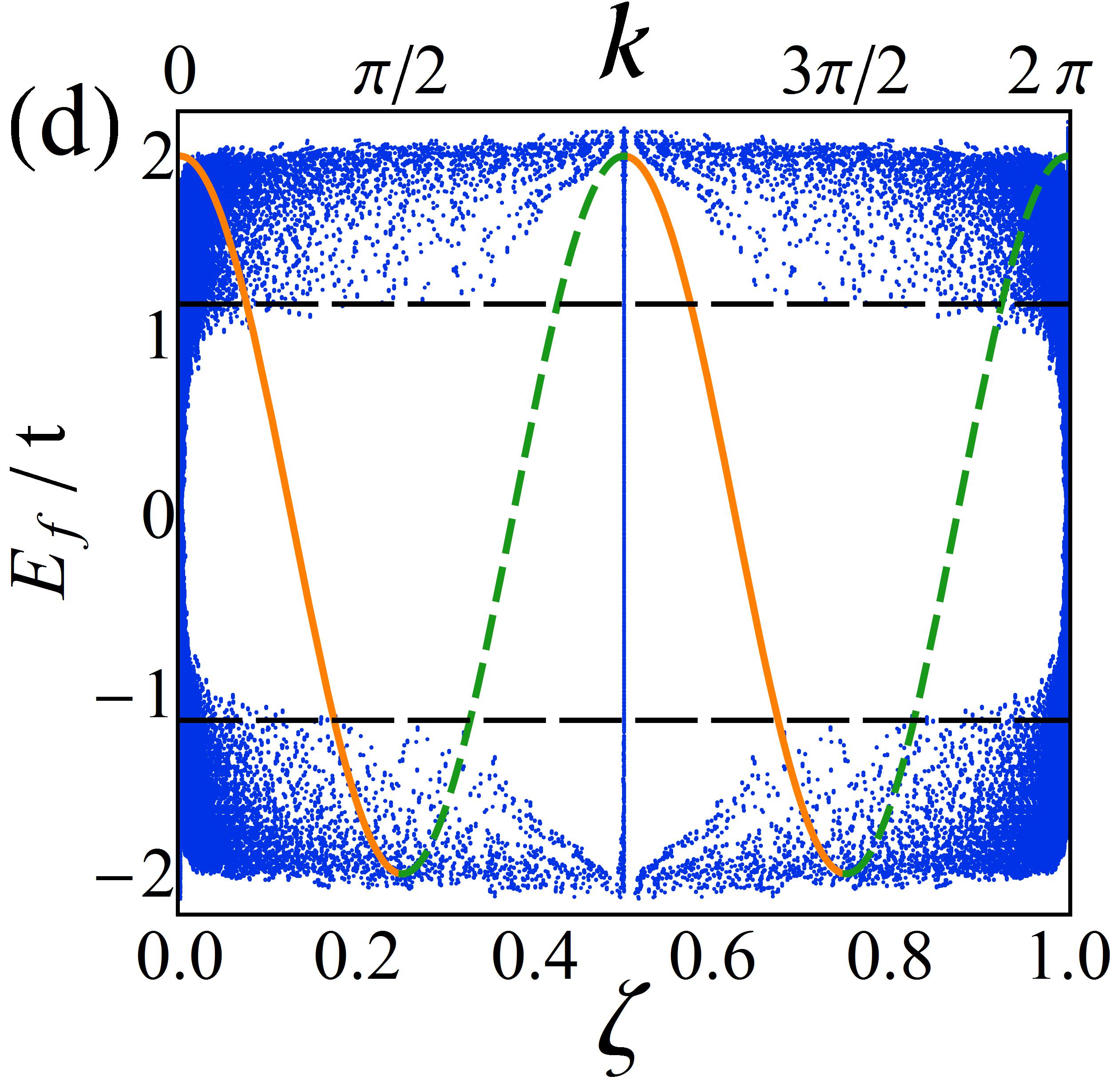}
\caption{(a) and (b): spatial and Fourier components of $\mathcal{V}$ respectively. (c) and (d): entanglement spectrum as a function of the Fermi level.  The dispersion relation is overlaid on the entanglement spectrum as a guide, with the restrictions on $\mathcal{A}$ schematically emphasized by the solid orange segments. The dashed green segments correspond to region $\mathcal{B}$, which is traced out. The black dashed lines delineate the regions over which one expects suppression of entanglement according to the zeroes of $\tilde{\mathcal{V}}_{\Delta k}$.}\label{EntVsFermiAlt}
\end{center}
\end{figure}

Finally, we consider an artificial model in order to verify that: (i) the presence of zeroes in the momentum-space matrix elements generically lead to resonances, regardless of the model-dependent shape that the disorder might have; (ii) the interpretation in terms of left and right-mover entanglement and momentum space partitioning is consistent.
To accomplish (i), we construct a new disorder distribution $\mathcal{V}_n$ such that $\tilde{\mathcal{V}}_{\Delta k}=0$ for all $\Delta k,$ except those in some finite ranges for which we set the magnitude to be a nonzero, constant value. Disorder is introduced by randomizing the phases of the scattering elements in momentum space.  An example profile of such a $\mathcal{V}_n$ and $\vert \tilde{\mathcal{V}}_{\Delta k}\vert$   are shown in Figs. \ref{EntVsFermiAlt}a,b  for a particular realization of the disorder.  To address (ii), we implement an exclusively next-nearest neighbor hopping model, so that the group velocity becomes $v_g(k)= -4t \sin 2k.$ In doing this, we consider two possible partitions of the Hilbert space: one choice corresponds to the restriction $k_{\mathcal{A}}\in[0,\pi]$, which is what we chose for the RDM, and the other takes into account the new location(s) of the left movers in momentum space, so that $k_{\mathcal{A}}\in[0,\pi/2]\cup[\pi,3\pi/2]$.  
The resulting entanglement spectrum versus Fermi energy, for both choices of partitions, is shown in Figs. \ref{EntVsFermiAlt}c,d. The dispersion relation is overlaid on the entanglement spectrum as a guide, with the restrictions on $\mathcal{A}$ schematically emphasized by the solid orange segments. The dashed green segments correspond to region $\mathcal{B}$, which is traced out. It is clear that the left/right mover cut provides the cleaner signature (since we are not getting spurious entanglement from left/left and right/right coupling) and that our interpretation of the delocalized states existing where the Fourier components of $\mathcal{V}_n$ are vanishing is correct.

We have shown that momentum space entanglement contains valuable information for the identification of localized and delocalized states in disordered free-fermion models. The next step is to generalize the method to higher dimensions to see if the entanglement signatures are as clear as in 1D. It is not obvious whether or not these results will easily generalize, but there will be some interesting features of momentum space entanglement in higher dimensional systems, and especially in topological phases with 1D edge states and extended states buried in a localized bulk state region. This method may also provide a nice way to characterize disordered interacting systems since it only requires a ground-state wavefunction and not the excited states. Momentum space cuts could be carried out efficiently for 1D interacting systems and may yield useful information.

\textit{Acknowledgements}
We would like to thank P. Phillips for several useful discussions and for introducing us to the random dimer model. TLH is supported by 
U.S. Department of Energy, Division of Materials Sciences under Award No. DE-FG02-07ER46453. 

\end{document}